\documentclass[12pt, letterpaper]{article}

\pdfoutput=1

\usepackage{arxiv}
\usepackage[utf8]{inputenc} % allow utf-8 input
\usepackage[T1]{fontenc}    % use 8-bit T1 fonts
\usepackage{amsmath,amsfonts,amssymb,bm}
\usepackage{mathtools}
\usepackage{mathptmx}
\usepackage{newtxtext}
\usepackage{newtxmath}
\usepackage{hyperref}
\usepackage{caption}
\usepackage{float}
\usepackage{xcolor}
\usepackage{multirow}
\usepackage{subcaption,booktabs}
\usepackage{rotating}
\usepackage{adjustbox}
\usepackage{graphicx}
\usepackage{array}
\usepackage{environ}
\usepackage[square,sort,comma,numbers]{natbib}
\title{Changepoint Detection: An Analysis of the Central England Temperature Series}

\author{
 Xueheng Shi \\
Department of Statistics,\\
University of California, Santa Cruz \\
 \And
 Claudie Beaulieu \\
Department of Ocean Sciences, \\
University of California, Santa Cruz\\
  \texttt{beaulieu@ucsc.edu} \\
  \And
Rebecca Killick \\
Department of Statistics,\\
 Lancaster University\\
  \And
Robert Lund \\
Department of Statistics,\\
University of California, Santa Cruz \\
}

\begin{document}

\maketitle

\begin{abstract}
This paper presents a statistical analysis of structural changes in the Central England temperature series, one of the longest surface temperature records available. A changepoint analysis is performed to detect abrupt changes, which can be regarded as a preliminary step before further analysis is conducted to identify the causes of the changes (e.g., artificial, human-induced or natural variability). Regression models with structural breaks, including mean and trend shifts, are fitted to the series and compared via two commonly used multiple changepoint penalized likelihood criteria that balance model fit quality (as measured by likelihood) against parsimony considerations. Our changepoint model fits\textcolor{black}{, with independent and short-memory errors,} are also compared with \textcolor{black}{a different class of models termed} long-memory models that have been previously used by other authors to describe persistence features in temperature series. In the end, the optimal model is judged to be one containing a changepoint in the late 1980s, with a transition to an intensified warming regime. This timing and warming conclusion is consistent across changepoint models compared in this analysis. The variability of the series is not found to be significantly changing, and shift features are judged to be more plausible than either short- or long-memory autocorrelations. \textcolor{black}{The final proposed model is one including trend-shifts (both intercept and slope parameters) with independent errors.} The analysis serves as a walk-through tutorial of different changepoint techniques, illustrating what can be statistically inferred.
\end{abstract}

\section{Introduction}
Climate time series often contain abrupt changes and other nonlinearities in their behavior. Changepoints are times of abrupt shifts in a series' characteristics, including means, trends, variances, and autocorrelations.  For examples, a sudden change from a cooling period (i.e., decreasing trend) to a warming period can be characterised by a changepoint in the trend; a sudden increase due to the relocation of a station may be characterised as a changepoint in the mean. Abrupt changes may be caused by changes in climate forcings, related to climate variability in the ocean and atmosphere, or induced by artificial changes in measurement procedures such as station relocations or instrumentation changes. 

It is crucial to know changepoint times in climate series, especially when assessing long-term trends, as their presence may grossly alter trend estimates, which impedes our understanding of external forcings and climate variability over the instrumental record \citep{Reeves_etal_2007, Beaulieu_2012_PTRS, Cahill_etal_2015, Beaulieu_Killick_2018}. Series with artificial changes merit adjustment via homogenization methods, as trends and extreme quantiles are more accurately estimated from homogenized data \citep{Lund_2017_Homogenization, Trewin_2020_Homogenization, Vincent_2020_Homogenization}. On average, approximately six station relocations or instrumentation changes occur over a century in a randomly selected US climate station \citep{Mitchell_1953, Menne_etal_2009}. As such, a changepoint analysis of a climate series is often a worthy initial exploratory endeavor.

Statistical methods to detect changepoints have rapidly evolved over the last few decades.  These include methods to detect a single shift in the series' mean \citep{Chernoff_Zacks_1964}, in its variance \citep{Hsu_1977}, or in a general linear regression model \citep{Quandt_1958, Robbins_Gallagher_Lund_JASA2016}.  In the climate literature, changepoint detection has most often been used to detect mean shifts. However, this may result in misinterpreting a long-term climate trend as a sequence of mean shifts that follows (approximates) the trend \citep{Beaulieu_Killick_2018}.

Much of the changepoint literature assumes independent and identically distributed model errors (termed white noise here). However, climate time series are often autocorrelated, inducing memory at time scales longer than the measurement frequency \citep{Hasselmann_1976}. This memory is often modeled as a first-order autoregressive (AR(1)) process in climate studies \citep{Reeves_etal_2007, Robbins_etal_2011, IPCCHartmann2013}.  In an AR(1) model, 
autocorrelation geometrically decays to zero with increasing time, representing one type of short-term memory. In the climate setting, it is important to allow autocorrelation and mean shift model features in tandem as both can inject similar run patterns into a climate series. An alternative is to use pre-whitening techniques that mitigate the effects of autocorrelation \citep{Robbins_etal_2011, Serinaldi_Kilsby_2016}. \cite{Beaulieu_Killick_2018},  \cite{Shi_etal_2021_comparison}, and \cite{Shi_Autocovariance_2021} show that changepoint inferences can be drastically wrong if autocorrelation in a series is ignored. The memory in climate series has also been modeled as a long-memory process, where autocorrelation decays as a power law \citep{Yuan_etal_2015}. Long-memory processes and changepoint models can be confused as they both have similar spectrums. Unfortunately, this ambiguity may lead to mislead inferences. \cite{Beaulieu_al_2020} discuss how to distinguish changepoints and long-memory in surface temperatures.

Multiple changepoints may be present in climate series.  Methods designed to detect a single changepoint have been applied iteratively to estimate multiple changepoint configurations through a process known as binary segmentation \citep{scott1974binaryseg, Rodionov_2004}. Binary segmentation is now known to perform poorly in multiple changepoint problems \citep{Shi_etal_2021_comparison} (see \cite{Piotr_WBS} for an interesting attempt to fix binary segmentation). Penalized likelihood methods, the approach taken here, were developed in \cite{Davis_etal_2006, Lu_etal_2010, Killick_etal_2012, Li_Lund_2012} and tend to perform better \citep{Shi_etal_2021_comparison}. Here, a likelihood, which measures the goodness of the statistical model fit, is balanced against a penalty that prevents fitting too many changepoints. Penalized likelihood methods can allow for autocorrelation. Bayesian approaches to the multiple changepoint problem also exist.  Most of these place some sort of prior distribution on the changepoint times, for instance a spike and slab prior (see \cite{Barry_Hartigan_1993, Chib_1998, Fearnhead_2006}; and \cite{Cappello_etal_2021} and the references within).  \cite{Li_Lund_2019} construct an informative prior on the changepoint times from the station's metadata record.  The references above are by no means exhaustive; indeed, the changepoint literature is vastly expanding.  

As most methodological statistics papers are not written with user comprehension in mind, the technical changepoint literature can seem impenetrable to non-statisticians, making it challenging to select an appropriate approach for the climate scientist.  Compounding difficulties, \cite{Lund_Reeves_2002} and \cite{Beaulieu_Killick_2018} show that spurious changepoint inferences easily occur when prominent data features (e.g. autocorrelation, long-term trend) are ignored --- the choice of model and method is critical in changepoint analyses.  Indeed, changepoint techniques can produce different results when the models and assumptions are only slightly changed.

The aim of this paper is to present, through an example, a comprehensive changepoint analysis of a climate series. To this end, we analyze the Central England temperature (CET) series by fitting different changepoint models capable of detecting shifts in trends. We also compare our changepoint fits with long-memory models.  Our focus is on penalized likelihood multiple changepoint techniques, enabling us to compare several models while preventing overestimation of the number of changepoints. We also discuss mean shift models and how they fit data containing a long-term trend such as the CET series.  Emphasis is placed on implementation and interpretation over the theoretical foundations of penalized likelihoods. Nonetheless, references to the formal statistical literature are provided.

The rest of this paper proceeds as follows. The CET series used here is introduced in the next section. Section 3 then provides some rudimentary background on changepoint models, describing the penalized likelihood methods used here. The next three sections present fits of various multiple changepoint models.  Results for each type of model motivate the subsequent fits.  Remarks about the optimal model are made in the final section along with concluding comments.

\section{The CET Series}

The CET time-series is perhaps the longest instrumental record of surface temperatures in the world, commencing in 1659 and spanning 362 years through 2020. The CET series is a benchmark for European climate studies, as it is sensitive to atmospheric variability in the North Atlantic \citep{Parker_etal_1992}. This record has been previously analyzed for long-term changes \citep{Plaut_etal_1995, Harvey_Mills_2003, Proietti_Hillebrand_2017}; however, to our knowledge, no detailed changepoint analysis of it has been previously conducted.  Changepoints are plausible in the CET record for several reasons.  First, artificial shifts near the record's onset may exist when data quality was lower \citep{Parker_etal_1992}. %Second, shifts associated with the Maunder minimum, the era circa 1645-1715 where sunspot activity was minimal and Europe was cooling, may be present \citep{Plaut_etal_1995}.  Finally, 
Furthermore, an increase in the pace of climate warming arising globally during the 1960s-1970s \citep{Beaulieu_Killick_2018, Cahill_etal_2015} may be present. The length of the CET record affords us the opportunity to explore a variety of temperature features.

\begin{figure}[t]
\caption{Station locations and annual average temperatures of Central England.} 
\centering
\noindent\includegraphics[scale=1.25]{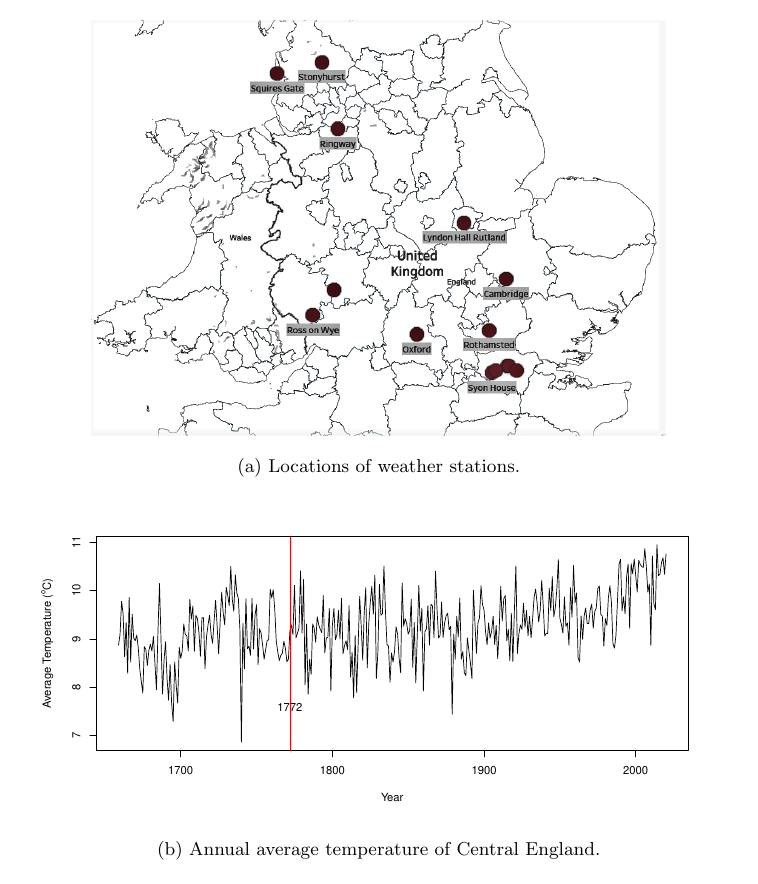}\\
\label{fig:CentralEnglandTemp}
\end{figure}

The CET series, available at \url{https://www.metoffice.gov.uk/hadobs/hadcet/}, was provided by the UK Met Office. Measurements commenced in 1659 and were mostly compiled by \cite{Manley_1953, Manley_1974} until 1973, then continued and updated to 1991 in \cite{Parker_etal_1992}. The series is now kept by the Hadley Centre, Met Office. The CET time series is an annual composite of 15 stations in the UK, located over a roughly triangular area bounded by Lancashire, London, and Bristol.  The series is thus representative of the climate of the English Midlands. The station locations used to form the composite series are depicted in the top graphic in Figure \ref{fig:CentralEnglandTemp}. The CET temperatures, presented in the bottom graphic of Figure \ref{fig:CentralEnglandTemp}, have been previously adjusted for inhomogeneities due to changes in measurement practices through time \citep{Manley_1953, Manley_1974, Parker_etal_1992}, and for urban warming since 1960 \citep{Parker_Horton_2005}. However, until 1722, available instrumental records used in the CET time series did not overlap. As such, non-instrumental weather diaries and the Utrecht instrumental series were used to adjust the CET series and fill the gaps \citep{Parker_etal_1992}. Between 1722 and 1760, there are no gaps in the composite record of all stations, but observations were generally collected in unheated rooms as opposed to outdoors. A few outdoor temperature measurements were collected and used to establish relationships between temperatures in unheated rooms and outdoors. These relationships were then used to adjust the CET time series \citep{Parker_etal_1992}. The daily CET time series starts in 1772, and has been used to update the monthly series (Parker et al., 1992). As such, some authors use only the data post-1772 for their analyses \citep{Proietti_Hillebrand_2017}. In this paper, we conduct a changepoint analysis on both the full CET time series (1659-2020) and the truncated series (1772-2020) that excludes the poorer quality data at the beginning of the record.

\section{Structural Change Models}

To explore structural changes in the CET series, a hierarchical changepoint analysis, gradually building on past findings, will be conducted.  Let $X_t$ denote the annual temperature observed at time $t$ and suppose that data from the years $1, \ldots, N$ are available. In general, a changepoint analysis partitions the series into $m+1$ distinct regimes, each regime having homogeneous characteristics.  The number of changepoints $m$ is unknown and needs to be estimated from the series. Let $\tau_i$ denote the $i$th changepoint time; boundary conditions take $\tau_0=0$ and $\tau_{m+1}=N$. 

All regression models in this paper have the time series regression form
\begin{equation}
\label{eq:regress}
X_t = f(t) + \epsilon_t, \qquad t = 1, 2, \ldots, N,
\end{equation}
where $f(t)=E[ X_t ]$ is the mean of the series at time $t$.  The structural form of $f$ will vary, generally containing location and/or trend parameters and their shifts; each model form will be discussed as we proceed.  The model errors $\{ \epsilon_t \}_{t=1}^N$ have zero mean and may be correlated in time. We work with AR(1) errors for simplicity, but more complex time series models are possible.  While it is important to allow for autocorrelation in annual data, the form of the correlation structure is typically not as crucial as its presence.  

The AR(1) difference equation governing the errors $\{ \epsilon_t \}$ is 
\begin{equation*}
\epsilon_t = \phi \epsilon_{t-1} + Z_t,
\label{e:ar1error}
\end{equation*}
where $\phi \in (-1,1)$ and $\{ Z_t \}$ is zero mean white noise (WN) with unknown variance $\sigma^2$.  Solutions to the AR(1) equation have exponentially decaying correlations:   $\mbox{Corr}(\epsilon_t, \epsilon_{t+h})=\phi^h$ for $h \geq 0$. Because the data are annually averaged, Gaussian distributed errors $\{ \epsilon_t \}$ are statistically realistic.  An implication of this is that future model likelihood functions will be Gaussian based.

Methods for handling multiple changepoint analyses without penalized likelihoods exist.  One popular technique is termed binary segmentation \citep{scott1974binaryseg}.  Binary segmentation works with any single changepoint technique, termed an at most one change (AMOC) method.  Many AMOC tests have been developed, including cumulative sums (CUSUM) \citep{page1954cusum}, likelihood ratios \citep{jandhyala2013lrt}, Chow tests \citep{chow1960}, and sum of squared CUSUM tests \citep{Shi_etal_2021_comparison}.  Binary segmentation first analyzes the entire series for a changepoint. If a changepoint is found, the series is split into subsegments about the identified changepoint time and the two subsegments are further scrutinized for additional changepoints.  The procedure is repeated iteratively until no subsegments are deemed to have changepoints.  While simple and computationally convenient, binary segmentation is one of the poorer performing multiple changepoint techniques \citep{Shi_etal_2021_comparison}, often being fooled by changepoints that occur close to one another or multiple shifts that move the series in opposite directions.  There have been attempts to fix binary segmentation --- see the wild binary segmentation and related methods in \cite{Piotr_WBS} and \cite{kirch2018mosum}. Unfortunately, these techniques typically assume independent model errors or are restricted to single parameter changes per regime (for example, mean shifts only). Perhaps worse, wild binary segmentation tends to overestimate changepoint numbers when they are in truth infrequent \citep{Lund_2020_WBS2}. 

To estimate the changepoint structure and model parameters from the data, penalized likelihood methods will be used.  Likelihood methods choose the model parameters that make seeing the observed data most likely; a penalty is imposed on the changepoint configuration to keep the fitted model parsimonious (from having too many changepoints).  Our penalized likelihoods have the following form
\begin{equation}
\label{eqn:penalized_fun}
-2 \log(L^*(m; \tau_1, \ldots, \tau_m)) + P(m; \tau_1, \ldots, \tau_m).
\end{equation}
The notation here is as follows:  $L^*(m;\tau_1, \ldots, \tau_m)$ is the optimal Gaussian likelihood that can be achieved from a model with $m$ changepoints that occur at the times $\tau_1, \ldots, \tau_m$.  Here, the data sample $X_1, X_2, \ldots, X_N$ is regarded as fixed.  To determine $L^*(m; \tau_1, \ldots, \tau_m)$, one must estimate all parameters in the mean function $f$ and the AR(1) model errors assuming that $m$ changepoints occur at the times $\tau_1, \ldots, \tau_m$.  This procedure will be discussed further below.  The quantity $P(m; \tau_1, \ldots, \tau_m)$ is the penalty for having a model with $m$ changepoints at the times $\tau_1, \ldots, \tau_m$.  As more and more changepoints are added to the model, the overall fit gets better ($-2\log(L^*)$ gets smaller); the penalty, which is positive and increases with the number of changepoints, prevents an overfitted model (one with too many changepoints).  

Many penalty structures have been proposed in the statistics and climate literature. These include the Akaike information criterion (AIC), the Bayesian information criterion (BIC), the modified Bayesian information criterion (mBIC), and Minimum description lengths (MDL).  We will use BIC and MDL here.  These two penalties were judged as "winners" in a recent changepoint detection comparison in \cite{Shi_etal_2021_comparison}. AIC penalties are not considered here because they often erroneously estimate an excessive number of changepoints \citep{Shi_etal_2021_comparison}. The BIC penalty for having $m$ changepoints at the times $\tau_1, \ldots, \tau_m$ is $m\log(N)$ and is proportional to the number of changepoints; additional parameters are penalized at the rate of $\log(N)$ per model parameter. Our penalized likelihood objective functions for structural changes are summarized in Table \ref{tab:all_penalized_lik_obj}.  The individual models will be explained in subsequent sections. The boxed quantities are the model penalties.  When $m=0$, penalties for any changepoint quantities are taken as zero since changepoint features are absent from the model.

When comparing models via BIC (or any other model selection criterion), one computes the BIC statistic for all fitted models and chooses the one with the smallest BIC score.  Differences between BIC values can give a sense of uncertainty between different model fits.  The ``posterior model probabilities" of 
\cite{Burnham_Anderson_2004} can further highlight differences.  Elaborating, we label the compared models as $g_i, (i=1, \ldots , R)$ and let $\Delta BIC_i$ denote the difference between the BIC score of model $g_i$ and the model having the smallest BIC score. The posterior model probabilities of \cite{Burnham_Anderson_2004} are
\begin{equation}
\label{eqn:BICweights}
p_i = \frac{exp(-\Delta BIC_i/2)}
{\sum_{r=1}^{R} {exp(-\Delta BIC_r/2)}}.
\end{equation}
Then $p_i$ is the inferred probability that model $g_i$ is the quasi-true model in the model set under a prior where all $R$ models are equally likely (prior probabilities are $1/R$ for each model). These BIC posterior model probabilities highlight uncertainties in our model comparisons.

In contrast to the BIC penalty, the MDL penalty is more complex in form, also accounting for the changepoint location times $\tau_1, \ldots, \tau_m$. The MDL penalty depends on the form of $f$ and is rooted in information theory, quantifying the computer memory needed to store the model (good fitting models use minimal space). MDL penalties have previously proven useful in changepoint detection \citep{Davis_etal_2006, Li_Lund_2012}). Posterior model probabilities are not available for the MDL information criterion. Other penalties used in the climate literature for changepoint problems include those in \cite{Caussinus_Mestre_2004}.  

\begin{table}
\caption{Penalized likelihoods. The boxed terms are the penalties, with the unboxed terms constituting $-2\log(L^*)$.  Here, $N$ denotes the length of series, $m$ the number of changepoints, $\tau_i$ is the time of the $i$th changepoint, and $\hat{\sigma}^2$ is the estimated white noise variance.}
\label{tab:all_penalized_lik_obj}
\begin{subtable}{1\textwidth}
\centering
\scalebox{1}{
\begin{tabular}{@{} c  l  @{}}
Criteria    &   Objective Function \\
\hline
BIC      & $N\log(\hat{\sigma}^2)+N+N\log(2\pi) + \boxed{(3m+4)\log(N)}$     \\
MDL & $N\log(\hat{\sigma}^2)+N+N\log(2\pi) +\boxed{2\log(N)+ 2\log(m) + 2\sum_{i=1}^{m+1} \log (\tau_i - \tau_{i-1}) + 2\sum_{i=2}^{m+1} \log(\tau_i)}$ \\
\end{tabular}
}
\caption{Penalized likelihoods for the trend shift model with AR$(1)$ errors}
\label{tab:trend_break_objfun_ar1}
\end{subtable}
\begin{subtable}{1\textwidth}
\centering
\scalebox{1}{
\begin{tabular}{@{} c  l  @{}}
Criteria    &   Objective Function \\
\hline
BIC      & $N\log(\hat{\sigma}^2)+N+N\log(2\pi) +\boxed{(3m+3)\log(N)}$     \\
MDL & $N\log(\hat{\sigma}^2)+N+N\log(2\pi) +\boxed{\log(N)+ 2\log(m) + 2\sum_{i=1}^{m+1} \log (\tau_i - \tau_{i-1}) + 2\sum_{i=2}^{m+1} \log(\tau_i)}$ \\
\end{tabular}
}
\caption{Penalized likelihoods for the trend shift model with white noise errors}
   \label{tab:trend_break_objfun_wn}
\end{subtable}
\begin{subtable}{1\textwidth}
\centering
\scalebox{1}{
\begin{tabular}{@{} c  l  @{}} 
Criteria    &   Objective function \\
\hline
BIC      & $N\log(\hat{\sigma}^2)+N+N\log(2\pi) + \boxed{(2m+4)\log(N)}$     \\
MDL & $N\log(\hat{\sigma}^2)+N+N\log(2\pi)+\boxed{3\log(N) + 2\log(m) + \sum_{i=1}^{m+1} \log (\tau_i - \tau_{i-1}) + 2\sum_{i=2}^{m+1} \log(\tau_i)}$ \\
\end{tabular}
}
\caption{Penalized likelihoods for the fixed slope mean shift with AR$(1)$ errors}
\label{tab:meanchange_constslope_objfun}
\end{subtable}
\begin{subtable}{1\textwidth}
\centering
\scalebox{1}{
\begin{tabular}{@{} c  l  @{}} 
Criteria    &   Objective function \\
\hline
BIC      & $N\log(\hat{\sigma}^2)+N+N\log(2\pi) + \boxed{(2m+1)\log(N)}$     \\
\end{tabular}
}
\caption{Penalized likelihoods for the Joinpin model with white noise errors}
\label{tab:joinpin_objfun}
\end{subtable}
\begin{subtable}{1\textwidth}
\centering
\scalebox{1}{
\begin{tabular}{@{} c  l  @{}} 
Criteria    &   Objective function \\
\hline
BIC      & $N\log(\hat{\sigma}^2)+N+N\log(2\pi) + \boxed{4\log(N)}$     \\
\end{tabular}
}
\caption{Penalized likelihoods for the long memory model with AR(1) errors. Minus $\log(N)$ for white noise errors.}
\label{tab:lm_objfun}
\end{subtable}
\end{table}

A drawback of penalized likelihood methods involves computation time.  There are $\binom{N-1}{m}$ distinct changepoint configurations having $m$ changepoints. Summing this over all $m$ shows that there are $2^{N-1}$ distinct changepoint configurations that need to be searched in an exhaustive optimization of a penalized likelihood, a daunting task for long time series.  As a solution, genetic algorithms (GA) will be used to optimize our penalized likelihoods. GAs are randomized search algorithms that mimic natural selection processes. In a genetic algorithm, an initial collection (generation) of changepoint configurations is randomly evolved towards ones with improved penalized likelihoods. Better fitting models are allowed priority in passing on their changepoints (genes) to children models of the next generation.  Occasionally, mutations (very different changepoint configurations) occur; this keeps the GA from converging to local minimums of the penalized likelihood.   Ultimately, the GA converges to a model with a very good penalized likelihood.  The natural selection mechanism in GAs make it unlikely to visit suboptimal changepoint configurations. While \cite{Li_Lund_2012} illustrate how to devise a GA in climate changepoint applications, generally available GAs have now become savvy enough to capably handle our needs.  The GA optimizations performed here use the \texttt{R} package \texttt{GA} \citep{scrucca2013ga}.

In contrast to GAs, binary segmentation is a greedy algorithm that often becomes trapped at a local penalized likelihood minimum.  \cite{Killick_etal_2012} and \cite{maidstone2017FPOP}, two rapid dynamic programming based multiple changepoint configuration optimizers, currently cannot handle our needs: \cite{maidstone2017FPOP} assumes independent model errors and \cite{Killick_etal_2012} assumes all parameters change at each changepoint time (including the AR(1) correlation parameter $\phi$ and error variance $\sigma^2$).  GAs are the only optimization method that reasonably handle all models considered in this paper.  

\section{Models fitted}
\subsection{Trend shift models}
We start our analysis with models having trends, as a long-term trend in the CET time series has been documented in previous studies \citep{kendon_2021, franzke_2012, Karoly_Stott_2006}. This model posits $f(\cdot)$ to have the piece-wise linear form
\begin{equation}
    \label{eqn:trend_break_mod}
    f(t) = 
    \begin{cases}
    \mu_1 + \beta_1 t, \qquad & 1\le t\le \tau_1,\\
    \mu_2 + \beta_2 t, \qquad & \tau_1+1 \le t\le \tau_2,\\
    \qquad \vdots \\
    \mu_{m+1} + \beta_{m+1} t, \qquad & \tau_m+1 \le t\le N,
    \end{cases}.
\end{equation}
More compactly, one can write $E[X_t]=f(t)=\mu_{r(t)}+\beta_{r(t)}t$, where $r(t) \in \{ 1,2, \ldots, m+1 \}$ denotes the regime being used at time $t$; for example, $r(t)=1$ for $1 \leq t \leq \tau_1$. 

The changepoint literature has focused primarily on detecting mean shifts; fewer studies have been dedicated to detecting trend shifts.  However, \cite{maidstone2017L0-penalty} present a dynamic programming approach that estimates trend shift configurations using a penalty based on absolute distances that is neither the MDL nor BIC.  Their $\{ \epsilon_t \}$ must be white noise (uncorrelated) with a zero mean and constant variance.  See \cite{bai1998change}, \cite{bai2003computation}, and their related \texttt{R} package \texttt{strucchange} by \cite{zeileis2015Rpackage} for more details.

The least squares estimators for the $i$th regime's parameters are computed from data in this regime only:
\begin{equation}
\label{eqn:trendshift}
\hat{\beta}_i=  \frac
{\sum_{t=\tau_{i-1}+1}^{\tau_{i} } (X_t-\bar{X}_i)(t-\bar{t}_i)} 
{\sum_{t=\tau_{i-1}+1}^{\tau_{i}} (t-\bar{t}_i)^2 }, \quad \hat{\mu}_i=\bar{X}_i-\hat{\beta}_i \bar{t}_i, \quad i=1,2,\ldots, m+1,
\end{equation}
where $\bar{X}_i =(\sum_{t=\tau_{i-1}+1}^{\tau_{i}}X_t)/(\tau_{i}-\tau_{i-1})$ and $\bar{t}_i=(\tau_{i}+\tau_{i-1}+1)/2$.   While these are not the exact maximum likelihood estimators in correlated settings, they are typically very close to them \citep{Lee_Lund_2012_JTSA}.  A detailed discussion of least squares versus maximum likelihood estimator differences for time series is contained in \cite{Lee_Lund_2012_JTSA}.

One next computes the detrended series via 
\begin{equation}
\label{e:resid}
D_t=X_t - \hat{f}(t)= X_t -(\hat{\mu}_{r(t)}+\hat{\beta}_{r(t)}t).
\end{equation}

The AR$(1)$ parameter is then estimated via
\begin{equation}
\label{e:phihat}
\hat{\phi} = \frac{\sum_{t=1}^{N-1} D_t D_{t+1}}
{\sum_{t=1}^N D_t^2}.
\end{equation}
One-step-ahead predictions of the time series are now computed by
\begin{equation}
\label{e:dhat}
\hat{D}_t = \hat{\phi} \hat{D}_{t-1}, \quad t \geq 2,
\end{equation}
with the start-up condition $\hat{D}_1=0$. The white noise variance in the AR(1) model is estimated as
\begin{equation}
\label{e:sig2hat}
\hat{\sigma}^2 = \frac{1}{N}\sum_{t=1}^N \hat{D}_t^2.
\end{equation}
Plugging $\hat{\mu}_k$, $\hat{\phi}$, and $\hat{\sigma}^2$ into the Gaussian likelihood (see \cite{Li_Lund_2012} for details) gives a negative Gaussian log-likelihood of 
\begin{equation}
\label{e:lkhopt}
-2 \log(L^*(m;\tau_1, \ldots, \tau_m)) = N \log (\hat{\sigma}^2) + 
\underbrace{N+N\log(2\pi).}_{\text{Constant}}
\end{equation}
The underbraced constant term above does not change over distinct changepoint configurations and can be neglected in the changepoint configuration comparisons.  The above equations show how to estimate model parameters and evaluate model likelihoods given the changepoint configuration; the optimal changepoint configuration is found by a GA search. The penalized likelihoods obtained with two different penalties, MDL and BIC, are presented in Table \ref{tab:all_penalized_lik_obj} for the various models used here.  Since regression lines are described by two parameters, all regimes are required to be at least three years long (so that fits in any single regime are not perfect). 

\begin{figure}[t]
\caption{Estimated CET trend shift structure. BIC and MDL flag the same changepoints in both the CET series ($1700,1739, 1988$, \textcolor{black}{red solid line}) and truncated CET ($1987$, \textcolor{black}{blue dashed line}) series when assuming either AR(1) or white noise errors.} 
\centering
\noindent\includegraphics[scale=0.62]{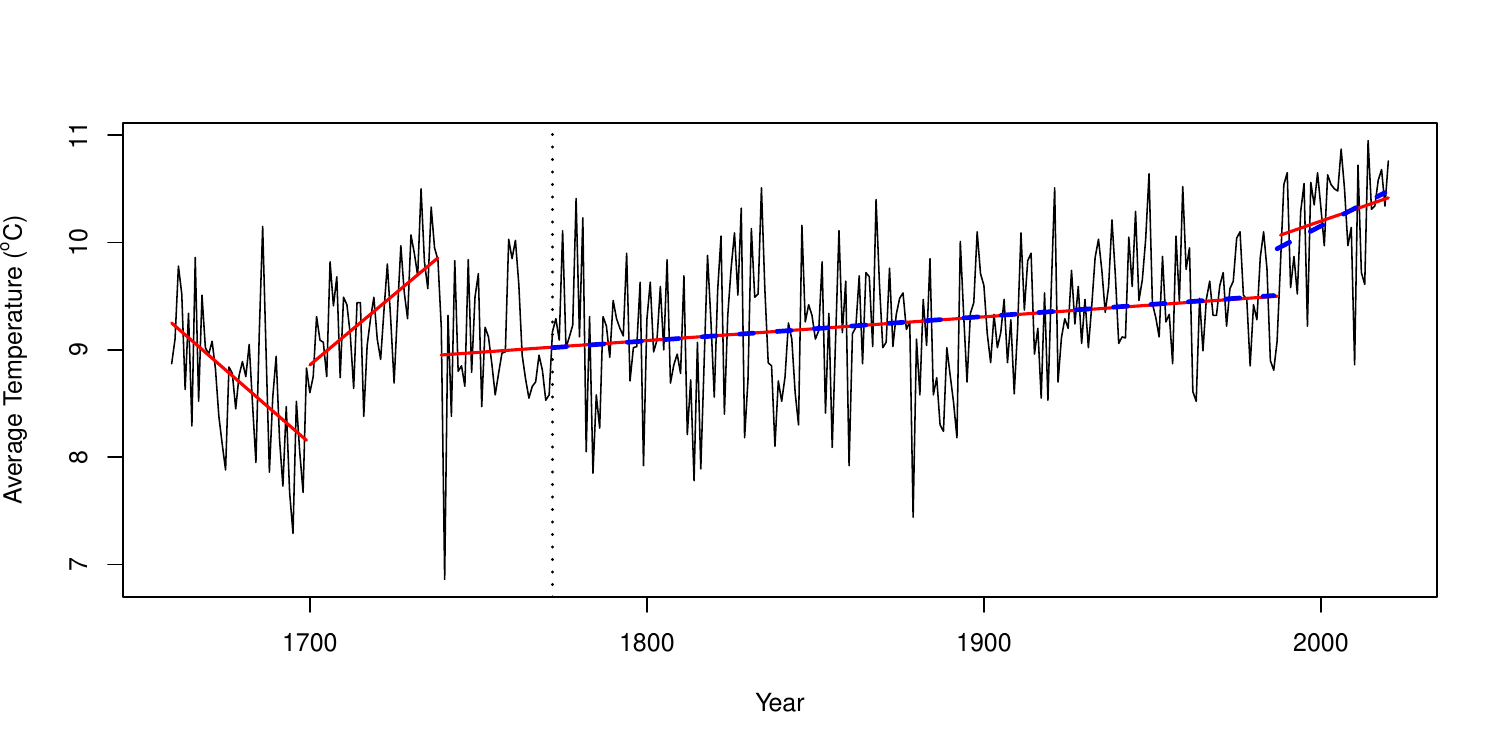}\\
\label{fig:TrendShift}
\end{figure}

On the full CET series, GA optimizations of the BIC and MDL penalized likelihoods estimate identical trend shift configurations, both flagging three breaks at the times $1700, 1739$, and $1988$ (Table \ref{tab:mod_fitting_result}).  This methodological agreement is convenient, but is not typical in changepoint analyses.  Figure \ref{fig:TrendShift} graphically depicts our model fit. Cooling occurs during the first 39 years, followed by an increasing-trend second regime, with subsequent shifts to two warming trend regimes.  The last regime, which starts in 1989, is warming with a trend of $1.1^\circ$C per century. When fitting trend shift models to CET series on post 1772 data only, we find a single changepoint in 1987 (Table \ref{tab:Truncatedmod_fitting_result}), which is consistent with our analysis on the full series.

In both cases, the AR(1) correlation estimate is very small ($\hat{\phi}=0.058$ for the full CET and $\hat{\phi}=0.073$ for the truncated), and is not significantly different from zero with standard time series tests \citep{Brockwell_Davis_1991}. When $\phi=0$, an AR(1) model reduces to white noise. This point is worth emphasizing: our model fits prefer the trend shift  structure over structures involving autocorrelated errors. This is an important point since positive autocorrelation and shifts can induce similar run patterns in series --- likelihood methods can decide which feature (or both) is statistically preferable. Should autocorrelation be neglected, one risks flagging spurious changepoints.  And while independent model errors is reasonable here, it may not hold in other applications, especially if monthly or daily data are used.

Other assumptions made on the model errors include normality and a constant variance in $X_t$ .  To assess normality, we apply a Shapiro-Wilk test to the model residuals.  This test does not reject normality (Tables \ref{tab:mod_fitting_result}- \ref{tab:Truncatedmod_fitting_result}) at any common levels of statistical significance. To investigate the constant variance assumption, we apply Leneve's test to the residuals.  This test does not find evidence of a changing variance in the residuals of the trend shifts models fitted to the CET series at any appreciable levels of statistical significance. Normality and constant variance assumptions in all future fitted models (Tables \ref{tab:mod_fitting_result} and \ref{tab:Truncatedmod_fitting_result} list these) is investigated --- these features are not rejected in any of the models compared here.

\subsection{A fixed slope mean shift model}
In some cases, it may be appropriate to constrain trends to be identical over all regimes \citep{Wang_JCcomments_2003}. This could be the case if artificial changes are expected. For example, a change of instrument may introduce an artificial shift in a time series, but will not necessarily alter the long-term trend in different regimes. A model with a common trend slope in all regimes \citep{Lu_Lund_2007_CJS} is 
\begin{equation}
    \label{eqn:meanbreak_fixedtrend_mod}
    f(t) = 
    \begin{cases}
    \mu_1 + \beta t, \qquad & 1\le t\le \tau_1,\\
    \mu_2 + \beta t, \qquad & \tau_1+1 \le t\le \tau_2,\\
    \qquad \vdots \\
    \mu_{m+1} + \beta t, \qquad & \tau_m+1 \le t\le N,
    \end{cases}.
\end{equation}

where $\beta$ is the trend slope, which is the same in all regimes.

In compact form, the model can be expressed as 
\begin{equation}
X_t = \mu_{r(t)} + \beta t + \epsilon_t,
\label{e:fixed_slope_meanshifts}
\end{equation}
where $\mu_{r(t)}$ is as in (\ref{eqn:trendshift}), and $\{ \epsilon_t \}$ is an AR$(1)$ process.

The ordinary least square estimators of $\beta$ and $\mu_1, \ldots, \mu_{m+1}$ have the explicit form
\begin{equation}
    \hat{\beta} = \frac{ \sum_{i=1}^{m+1} \sum_{\tau_{i-1}+1}^{\tau_{i}}(X_t-\bar{X}_i)(t-\bar{t}_i)}{ \sum_{i=1}^{m+1} \sum_{t=\tau_{i-1}+1}^{\tau_{i}}(t-\bar{t}_i)^2 }, \qquad \hat{\mu}_i = \bar{X}_i - \hat{\beta} \bar{t}_i, \qquad i=1,2,\ldots,m+1,
    \label{e:fixed_slope}
\end{equation}
where $\bar{X}_i$ and $\bar{t}_i$ are as before. These are again very close to the maximum likelihood estimators \citep{Lee_Lund_2012_JTSA}. The BIC and MDL penalties are listed in Table \ref{tab:all_penalized_lik_obj}.

\begin{figure}[t]
\caption{The estimated CET trend shift structure for the full \textcolor{black}{(red solid line)} and truncated CET \textcolor{black}{(blue dashed line)} series when a constant regime trend slope is imposed.  Both BIC and MDL flag a single changepoint in 1988 for the full series and 1990 for the truncated series.}
\centering
\noindent\includegraphics[scale=0.64]{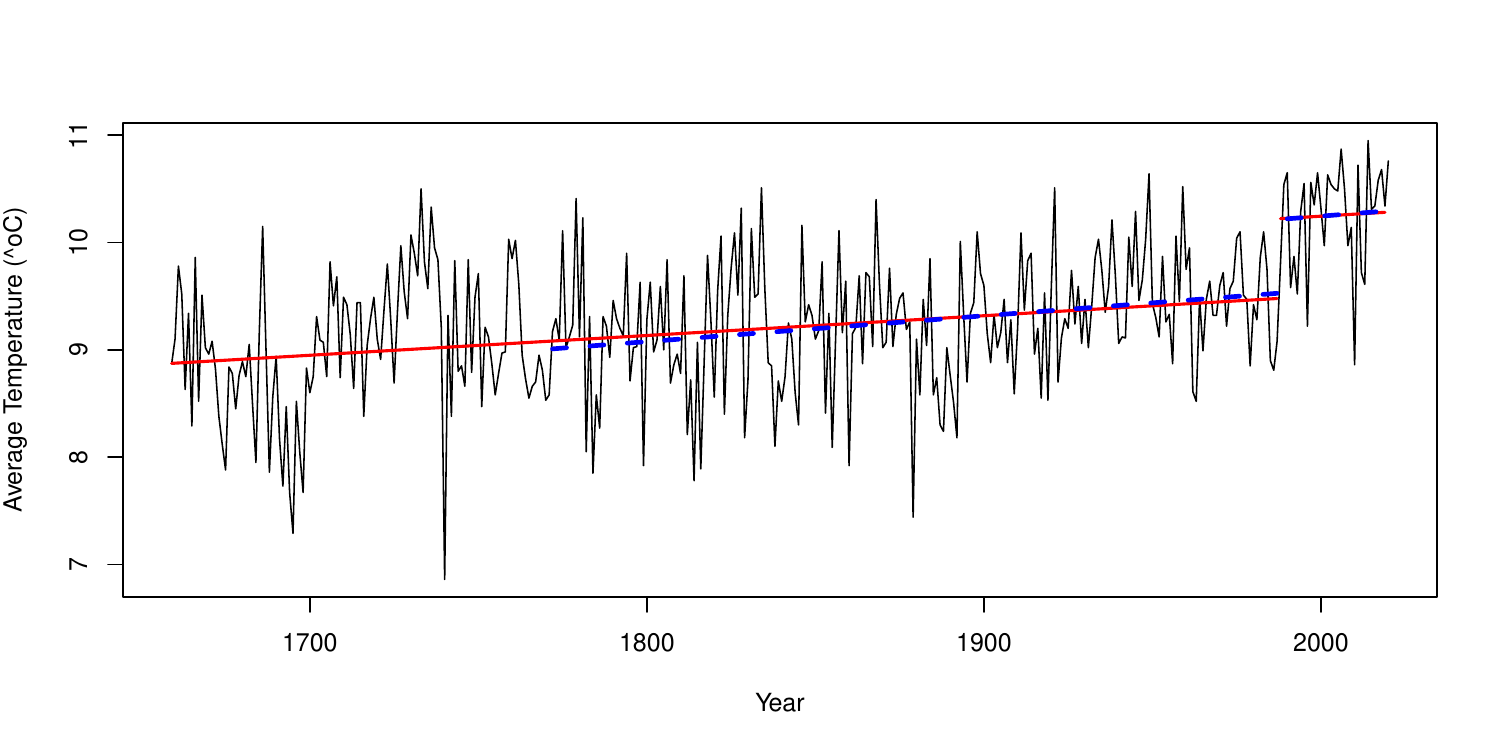}\\
\label{fig:MeanCpt_FixedTrend_AR1}
\end{figure}

A GA was used to estimate this configuration, which is plotted against the data in Figure \ref{fig:MeanCpt_FixedTrend_AR1}. For the full CET series, both BIC and MDL flag a single mean shift in 1988, while the single detected shift moves to 1990 in the truncated series (post 1772). Fewer changepoints are detected in this model than with the trend shift models of the previous section, but the time of the single change detected here is consistent with the last changepoint found in the trend shifts models. Since the BIC and MDL penalized likelihoods in Tables \ref{tab:mod_fitting_result} and \ref{tab:Truncatedmod_fitting_result} are larger for the constant slope model than for the regime-varying trend slope model, the inference is that regime-varying slopes are preferable. 

\subsection{Joinpin models}
There is debate over whether trend models should impose continuity in $E[X_t]$ at the changepoint times in temperature series \citep{Rahmstorf_2017}. These so-called joinpin models require $E[X_t]=f(t)$ to be continuous in time $t$.  Here, we compare a joinpin model to the trend shifts and fixed slope mean shift models fitted in the previous sections. Unfortunately, it is not clear what an appropriate MDL penalty is for this case, nor does this seem to be an easy matter to rectify; hence, we proceed with BIC penalties only.

\begin{figure}[t]
\caption{Estimated CET joinpin shift structure \textcolor{black}{for full (red solid line) and truncated (blue dashed line) series}.  BIC flags one shift in 1973 in the full series and and none for the truncated series.} 
\centering
\noindent\includegraphics[scale=0.6]{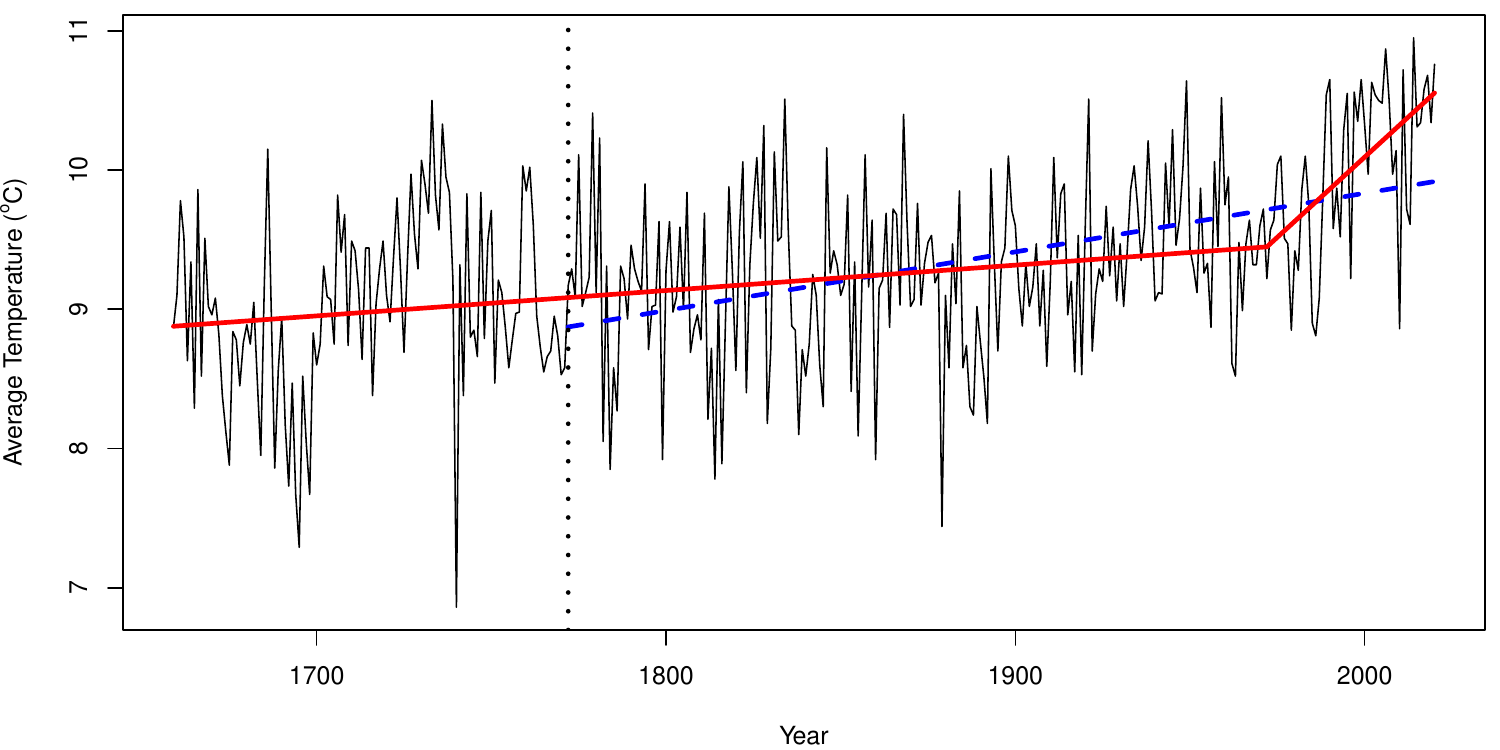}\\ 
\label{fig:JoinPinFullTruncated}
\end{figure}

To fit a joinpin model, the package in \cite{maidstone2017L0-penalty} was used. We fit the same model as \eqref{eqn:trend_break_mod}, but with additional constraints to force continuity at the changepoint time(s). A simple way to enforce this continuity is to view the slopes as determined from $E[X_t]$ at the start and end of each regime.  This enforces continuity within a simple form foregoing additional constraints.  This formulation fits the model
\begin{align}
  X_t=\gamma_{\tau_i}+ \frac{\gamma_{\tau_{i+1}}-\gamma_{\tau_i} }{\tau_{i+1}-\tau_i}(t-\tau_i) + \epsilon_t,
\end{align}
where $\gamma_i$ is the value of the mean at time $i$.  This formulation is equivalent to \eqref{eqn:trend_break_mod} with an additional continuity constraint at the changepoint locations. Based on \cite{maidstone2017L0-penalty}, the BIC for the joinpin model is 
\begin{equation}
\mbox{BIC} = N \log(\hat{\sigma}^2)+N
+ N\log(2\pi) 
+ (2m+1)\log(N), 
\end{equation}
where 
\[
\hat{\sigma}^2=\frac{1}{N}\sum_{i=1}^{m+1} \sum_{t=\tau_{i}}^{\tau_{i+1}} 
\left[
X_t -\frac{\gamma_{\tau_{i+1}}-\gamma_{\tau_i} }{\tau_{i+1}-\tau_i}(t-\tau_i) \right]^2.
\]
In the formulation of \cite{maidstone2017L0-penalty}, the white noise variance is fixed and needs to be estimated.  While median absolute deviations could be used for this purpose, we instead use the estimated error variance of $0.29$ (Table
\ref{tab:mod_fitting_result}), taken from the discontinuous model fits and BIC penalties of the last section,  This fit assumes IID errors, which seems plausible given the results of the previous sections.  The fitted model flags a single changepoint in 1973 in the full CET series and none in the truncated series; see Tables \ref{tab:mod_fitting_result}-\ref{tab:Truncatedmod_fitting_result} and Figure \ref{fig:JoinPinFullTruncated}. These fits are stable against changes from 0.29 in the white noise variance.  Compared to our previous model fits, the joinpin model has a much higher BIC than the trend shift and fixed slope mean shifts models (Tables \ref{tab:mod_fitting_result}-\ref{tab:Truncatedmod_fitting_result}).  As such, joinpin models do not appear to be competitive.

While a changepoint seems plausible towards the end of the record due to an  increased warming rate, the joinpin fit to the earliest data is poor, similar to the fixed slope mean shifts model.  This is graphically evident in the Figure \ref{fig:JoinPinFullTruncated} fits, but is also reflected by the higher BIC scores in Tables \ref{tab:mod_fitting_result}-\ref{tab:Truncatedmod_fitting_result}. A joinpin model should be used when a discontinuous mean function is unlikely or physically implausible. With the CET series, it is not evident whether the estimated mean function should be continuous or discontinuous. Elaborating, for series containing ``only a single station", mean discontinuities are physically expected. However, when more and more station records are averaged into a composite record, mean function discontinuities are reduced, becoming less pronounced with an increasing number of stations. Should a discontinuous mean function be deemed possible, a trend shift model provides greater flexibility since it can simultaneously approximate a joinpin continuous structure as well as discontinuous shifts \citep{Beaulieu_Killick_2018}. 

\subsection{Long-memory models}

A body of climate literature argues that climate time series exhibit long-memory, where the series' autocorrelation decays slowly in lag, often via a power law \citep{Yuan_etal_2015, Blender_Fraedrich_2003,franzke_2012}. Long-memory correlation and changepoint features can inject similar run properties into a climate series, which is appreciated in the statistical and econometric literatures  \citep{DIEBOLD2001, GRANGER2004, Mills_2007,yau_davis_2012}. The daily CET series may exhibit long-memory \citep{SYROKA2002,franzke_2012}.

To compare our changepoint models to a long-memory model, we fit an autoregressive fractionally integrated moving-average (ARFIMA) model to the CET series. In particular, ARFIMA models with no moving-average component, an integration parameter $d$ with $0 < d < 0.5$, and an autoregressive component of orders zero and one, are considered. The AR(1) long-memory model is characterized as
\begin{align}
    X_t=(1-B)^d (1-\phi B)^{-1}\epsilon_t,
\end{align}
where $B$ is the backshift operator applied to $X_t$.

To fit ARFIMA models, the \texttt{R} package \texttt{fracdiff} \citep{R:fracdiff} was used.  A BIC penalty was calculated and is listed in Table \ref{tab:all_penalized_lik_obj}.  An MDL penalty is not informative since this model does not have any changepoints. Long-memory model fits to the full and truncated CET series are described in Tables \ref{tab:mod_fitting_result}-\ref{tab:Truncatedmod_fitting_result}). The long-memory models have the largest BIC score among all models compared on the full CET time series. On the truncated series, they are also amongst the least plausible, although joinpin models have higher BIC scores. These results suggest that changepoints, rather than long-memory, are more plausible in the CET series. For additional evidence that changepoints are preferred over long-memory features, we applied the time varying wavelet spectrum methods in \cite{Norwood_Killick_2018} to the CET series. These methods were used on surface temperatures in \cite{Beaulieu_al_2020} and shown to discriminate changepoint and long-memory models well in long series. The results confirm that a changepoint model is more appropriate than a long-memory model. The fitted model of autoregressive order zero was also preferred to the fitted model of order one, reinforcing that correlation aspects in the CET series are minimal.

\subsection{Model selection uncertainty}
Among the six models compared, the trend shift model with white noise is judged the most plausible, as suggested by both BIC and MDL scores. The BIC posterior probabilities for all models fitted above are presented in Table \ref{tab:BICweights}. For the full series, the model probability for the trend shift model with white noise is 0.64, followed by the joinpin model with probability 0.12 and the trend shift model with AR(1) errors with probability  0.11. The three other models all have a posterior probability of 0.05 or less. This highlights the uncertainty in the model selected, although the trend shifts models with AR(1) and white noise errors are very similar (the autocorrelation estimated in the AR(1) model is small and both configurations identify the same shifts). As for the joinpin model, the fit at the start of the record seems poor.

Moving to the truncated series, the trend shift model with white noise has a posterior probability of 0.68. The next most plausible models are the fixed slope mean shift model with AR(1) errors and the trend shift model with AR(1) errors, having posterior probabilities of 0.1 and 0.09, respectively (Table \ref{tab:BICweights}). These models are similar in that estimated changepoint times are very close, giving further evidence for a shift in the late 1980s. However, this suggests that a fixed slope model should not be entirely discarded. Unlike results for the full CET series, the joinpin model ranks very low (0.02) on the truncated CET series. This is not surprising given that no changepoint is detected under the joinpin model in the truncated series (Figure \ref{fig:JoinPinFullTruncated}).

\section{Trends vs Mean Shifts}
The simplest changepoint analysis is arguably that of mean shifts. This is the most common model in the changepoint literature and has been widely used to analyze climate series. While this structure is inappropriate for series having trends (such as the CET analyzed here), we include this model here for comparative purposes. The mean shifts model posits $f(\cdot)$ to have form
\begin{align}
f(t)=\begin{cases}
     \mu_1, \quad & 1 \le t \leq \tau_1,\\
     \mu_2, \quad & \tau_1+1 \le t \leq \tau_2,\\
     \;  \quad \vdots  \\
     \mu_{m+1}, \quad &\tau_{m}+1 \le t \leq N.
     \end{cases}
     \label{e:meanshifts}
\end{align}
The model's mean structure is compactly written as $f(t)=E[X_t]=\mu_{r(t)}$, where $r(t) \in \{ 1,2, \ldots, m+1 \}$ denotes the regime being used at time $t$; for example, $r(t)=1$ for $1 \leq t \leq \tau_1$. 

Given $m$ and the changepoint times $\tau_1, \ldots, \tau_m$, mean parameters are first estimated via segment averages:
\begin{equation}
\hat{\mu}_i = \frac{1}{\tau_i - \tau_{i-1}}\sum_{t=\tau_{i-1}+1}^{\tau_i} X_t,\qquad i=1,2, \ldots, m+1.  
\label{e:segmeans}
\end{equation}
While sample means are not the exact maximum likelihood estimators of the mean parameters for correlated series, they are typically very close and are easy to compute (unlike maximum likelihood estimators). Next, the regime-wise mean estimated in (\ref{e:segmeans}) is subtracted from the series by computing $D_t=X_t-\hat{f}(t)=X_t -\hat{\mu}_{r(t)}$. The variance $\hat{\sigma}^2$ is then estimated as in (\ref{e:sig2hat}). We do not fit this model with AR(1) errors based on the results from the previous sections. The BIC and MDL penalized likelihoods for this model are
\begin{equation}
\mbox{BIC} = N\log(\hat{\sigma}^2)+N+N\log(2\pi) +(3m+3)\log(N);
\end{equation}
\begin{equation}
\mbox{MDL} = N\log(\hat{\sigma}^2)+N+N\log(2\pi) +\log(N)+ 2\log(m) + 2\sum_{i=1}^{m+1} \log (\tau_i - \tau_{i-1}) + 2\sum_{i=2}^{m+1} \log(\tau_i).
\end{equation}

\begin{figure}[t]
\caption{The estimated CET mean shift structure \textcolor{black}{for full (red solid line) and truncated (blue dashed line) series}. BIC and MDL detect the same changepoints for both the CET and truncated CET series assuming white noise errors.}
\centering
\noindent\includegraphics[scale=0.62]{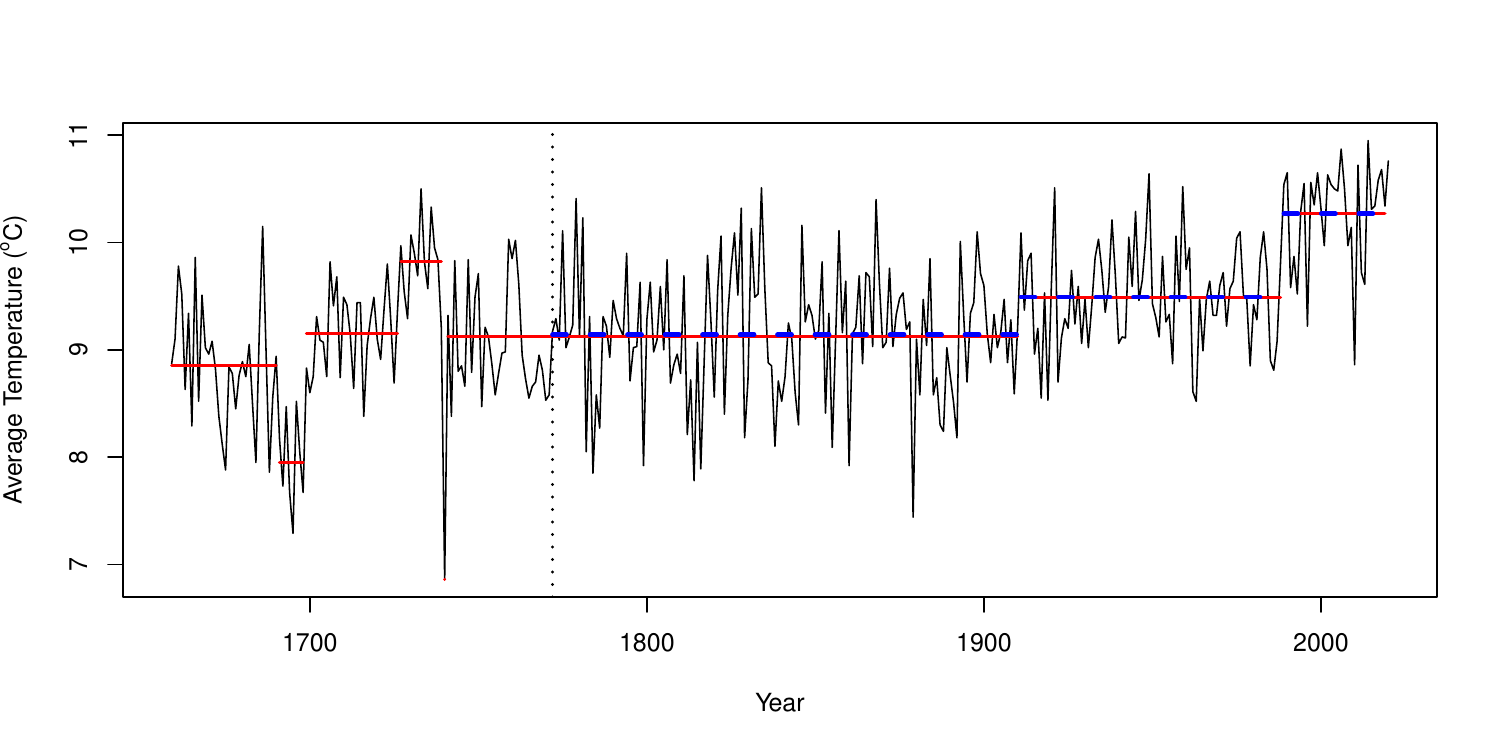}\\
\label{fig: Meanshifts-WN}
\end{figure}

We discuss only results on the full series here, but conclusions are consistent (i.e., the same changepoints are detected post 1772) if we repeat the analysis on the truncated series only. Fitting this model, seven changepoints are flagged with both MDL and BIC (Figure \ref{fig: Meanshifts-WN}).

Both penalties pinpoint 1989 as a changepoint time, which is consistent with results of the previous section. Here, MDL and BIC both deem the ``cold year" in 1740 an outlier, bracketing this time by two changepoints.  Because MDL methods are based on information theory \citep{Rissanen_1978} and not large sample statistical asymptotics, they often flag outliers. Shifts are more frequent at the beginning of the record, perhaps suggesting that the data during these times is less reliable.  Evident in the fits is that the last three regimes act to move the series higher in a "staircase", which is expected for a series experiencing a long-term warming trend (Figure \ref{fig: Meanshifts-WN}).

\begin{figure}[t]
\caption{Histogram of differences in BIC scores between the trend and mean-shift models. The correct model is the trend-shift model; however, BIC selects the mean-shift model the majority of the time.}
\centering
\noindent\includegraphics[width=0.8\textwidth]{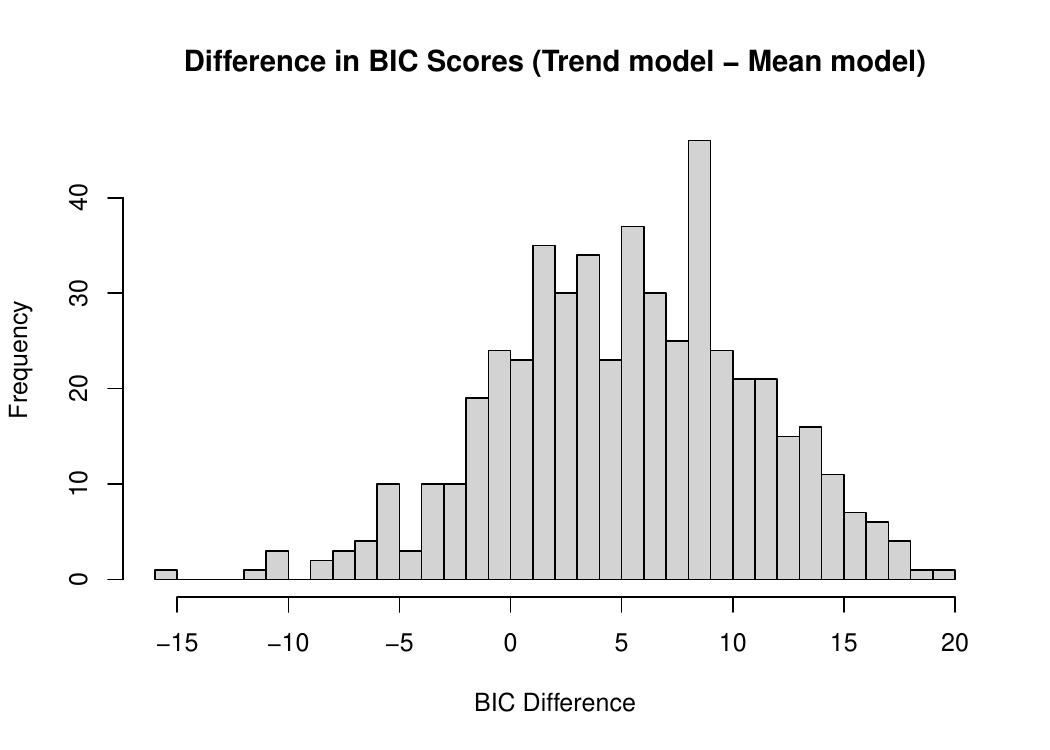}\\
\label{fig: simresults}
\end{figure}

The BIC and MDL scores obtained on the full CET series are 648.17 and 656.09, respectively. Should this model be included in our main comparison, one would still prefer the trend shift model should the MDL penalty be used to make conclusions. However, the BIC mean shift score is smaller than the BIC trend shift score in the previous section, indicating preference for the mean shift model.  A model containing only mean shifts will flag a sequence of shifts in an attempt to follow a long-term trend should the data have a trend and it not be included in the model. If the trend is not steep, as is the case here, it is especially challenging to distinguish between trends and mean shifts. To illustrate this, we conducted a simulation study where 500 synthetic series with the same trend magnitude and variability (as estimated in the truncated CET time series over 1772-2020) were generated. The mean shifts plus white noise and trend shifts plus white noise models were fitted to each series. In only $18\%$ of the synthetic series, the correct model with a long-term trend was selected by BIC. Figure \ref{fig: simresults} presents a histogram of the difference between the two fitted models' BIC scores, further demonstrating the bias BIC has for the erroneous mean shifts model. Should there be any suspicion about a trend or ``staircase feature" in the record, we recommend using techniques that incorporate trends, as done here. 

\section{Comments, Conclusions, and Discussion}
This study compared and contrasted several common changepoint model fits for data containing trends, as well as a long-memory autocovariance model, to the CET time series. To our knowledge, this is the first time a detailed changepoint analysis has been conducted on this long record. Starting with a trend shift model, several different changepoint structures were fitted, illustrating the techniques and salient points of changepoint analyses.

Tables \ref{tab:mod_fitting_result}-\ref{tab:Truncatedmod_fitting_result} present the log-likelihood, BIC, and MDL scores of all model fits.  Depending on the model configuration, we detect either three changepoints (trend shifts models) or one changepoint (fixed slope mean shifts and joinpin models) in the full series. This changepoint count discrepancy traces to the large variations in the series during roughly the first century of the record.

Most models agree on a change to a rapidly warming regime circa 1988, except for the joinpin model (this is also true for the truncated series). Among all fitted models, the optimal one has trend shifts in 1700, 1739, and 1988 (full series), and one in 1988 (truncated series). Table \ref{tab:best_fitting_mod} provides estimates of the best fitting model's intercept and slope parameters by regime. While the best fitting model is the trend shifts model, other models are also plausible (Table \ref{tab:BICweights}). Models with higher posterior probabilities tend to be consistent in their flagged changepoint times, but highlight that a fixed slope model (as opposed to the varying slopes in the trend shifts models) may be plausible. Long-memory models yield the highest BIC scores, and are less plausible than all other models compared. The results of the full and truncated CET series are consistent, showing that our post 1772 changepoint inferences are not overly sensitive to inclusion of the first century of the series. 

Having both BIC and MDL penalties agree on the model type and changepoint configuration adds robustness to our conclusions, suggesting that the fitted segmentations are stable. According to \cite{lavielle2005}, changepoint segmentations that are stable over a range of penalty values should be preferred. Overall, models with shifts were deemed preferable to models having autocorrelated errors. 

While our aim is not necessarily directed to the causes of the detected shifts, we provide some interpretations here. Shifts flagged during the first century of the record are likely due to inferior data quality over this early period \citep{Proietti_Hillebrand_2017}. Due to lack of overlapping instrumentation coverage before 1722, non-instrumental weather diaries were used to adjust the series \citep{Parker_etal_1992}. Observations were generally collected in unheated rooms until 1760, and adjusted by calibrating indoor and outdoor observations later \citep{Parker_etal_1992}. Even with the most careful adjustments, one cannot guarantee that all biases were removed from the data. Some authors omit the first century of data altogether due to this issue \citep{Proietti_Hillebrand_2017}. 

The trend shifts model on the earlier part of the data detects two changepoints in 1700 and 1739, characterizing a steep cooling trend followed by a warming trend. The mean shifts model fitted on the earlier part of the data flags multiple changes (1691, 1699, 1727, 1740, 1741), calling for a closer examination of the earlier part of the record. In data with inhomogeneities, BIC penalties favor mean shift models over trend shift models, even if the trend shifts model is truth.  A mean shift model characterizes a warming trend as a staircase of increasing steps. This issue can be troublesome if the trend in the data is weak, as demonstrated in our simulation study (see Figure \ref{fig: simresults}). %Although less likely, other potential explanations for the early shifts include a response to the Maunder minimum in the sunspot cycle  \citep{Plaut_etal_1995}. 
%Any volcanic activity at that time could further exacerbate the observed cooling as a response to solar irradiance forcing.

The changepoint flagged in 1988 (from multiple models and in both  the full and truncated CET series) is not surprising given the warming seen on the global level in the 1960/70s in a range of surface temperature records, as discussed in studies using both trend shift and joinpin models \citep{Cahill_etal_2015, Beaulieu_Killick_2018, Rahmstorf_2017, Ruggieri_2012}. While the more recent part of the CET series is considered more reliable and has been adjusted for inhomogeneities, we cannot entirely discard issues in this era either. Overall, it is possible that a combination of natural and artificial causes contribute to shifts in the CET series.

To further rule out artificial changes, one could subtract all $\binom{15}{2} = 105$ pairs of series from one another and examine these differences for changepoints. Then, one can distinguish artificially caused changepoints from those due to natural climate change and variability. See \cite{Menne_etal_2009} for more details on this procedure. \textcolor{black}{Artificial changes can then be corrected before long-term trends are analyzed. Changes that are not considered artificial can be further investigated through an attribution study \citep{IPCCHartmann2013}.} 

Residual analyses were conducted to ensure that the underlying assumptions of the model were met. With the CET series, residuals of the trend shift model fit were judged to be uncorrelated (white noise). However, climate time series often exhibit autocorrelation that should be taken into account.  We stress the importance of verifying the underlying assumptions in any changepoint model. Indeed, neglecting positive autocorrelation raises the risk of detecting spurious shifts.  Also, the series' autocorrelation may be more complex than an AR(1) process and may itself contain shifts \citep{Beaulieu_2012_PTRS, Beaulieu_Killick_2018}. Some climate series may also contain long-memory autocorrelations \citep{Vyushin_al_2012}. An additional challenge lies with the ambiguity between long-memory and changepoint models:  both features can produce series with similar run structures.  Because of this, a long-memory model was included as part of our comparison. We found that the CET time series is best represented by a multiple trend shift changepoint structure and not a long-memory model.  Such a comparison is not possible for all climate series since lengthy records are required to analyze long-memory series \citep{Beaulieu_al_2020}. The CET time series, which is the longest publicly available surface temperature series, enables this comparison. Other assumptions that were made include constant variance temperatures and normally distributed observations. Both assumptions cannot be rejected in any models fitted (Tables \ref{tab:mod_fitting_result}-\ref{tab:Truncatedmod_fitting_result}).

Model selection based on a criteria does not guarantee that the selected model is "truth". All models are an approximation of reality and multiple models can plausibly represent the data. To quantify this, one can calculate posterior model probabilities with BIC that each fitted model is the "quasi-truth". This assumes that all models included in the comparison have the same prior weight, which may not be reasonable. One must also note that this measure is relative to the models included in the comparison, and does not reflect the uncertainty that the "true" model may not be part of the model set. Similarly, uncertainty in the total number of changepoints and their individual occurrence times is a difficult statistics problem.  Bayesian methods, which were not considered here, can in principle place uncertainty margins on the number of changepoints and their locations.  When several distinct models have similar penalized likelihood scores, inferences about the number of changepoints are likely to be less reliable.  Recent statistics work is now studying this issue \citep{Li_Lund_2019, Cappello_etal_2021}.

\textcolor{black}{Ultimately the choice of "best model" should be arrived at from a judgment made by the researcher(s) based on objective  statistical metrics, such as presented in this work, combined with understanding of the data recording practices and physics of the natural system.}

\begin{table}
\centering
\caption{Model fitting results. Here, $\hat{\sigma}^2$ denotes the estimated variance of the white noise (* is assumed rather than estimated). Bolded values are the smallest penalized score. All model residuals have been checked for normality (Shapiro-Wilk's \& Kolmogorov-Smirnov test) and constant variance (Levene's test).}
\scalebox{0.7}{
\begin{tabular}{ccc|ccc }
\hline
Model %\tablefootnote{Residuals of all models fitted are normally distributed (examined by Shapiro-Wilk test)}
%\tablefootnote{ Residuals of all models fitted have a constant variance (examined by variance changepoint).}
& Penalty & Flagged Changepoints &$\hat{\sigma}^2$ & Log-likelihood  & Penalized Score  \\ \hline
\multirow{2}{*}{Trend shifts+AR(1)} & BIC &1700,1739,1988 &0.290  &-288.80 &654.19 \\
& MDL &1700,1739,1988 &0.290 & -288.80  &656.52 \\ 
\hline
\multirow{2}{*}{Trend shifts+WN} & BIC & 1700,1739, 1988 &0.291  & -290.02 & \textbf{650.74} \\
& MDL &1700,1739, 1988  &0.291 & -290.02  & \textbf{653.07}\\ 
\hline
\multirow{2}{*}{Fixed slope mean shift+AR(1)} & BIC & 1988 &0.325 & -310.11 &655.79 \\
& MDL  &1988  &0.325 &-310.11  &658.93\\
\hline
Joinpin & BIC &1973 &0.291* & -321.19 & 654.17\\
\hline
Long-memory+AR(1) & BIC & - &0.579 & -316.59 & 656.75\\
\hline
Long-memory & BIC & - &0.584 & -319.31 & 655.93\\
\hline
\end{tabular}
\label{tab:mod_fitting_result}
}
\end{table}

\begin{table}
\centering
\caption{Model fitting results based on truncated CET series. Here, $\hat{\sigma}^2$ denotes the estimated variance of the white noise (* is assumed rather than estimated). Bolded values are the smallest penalized score. All model residuals have been checked for normality (Shapiro-Wilk's \& Kolmogorov-Smirnov test) and constant variance (Levene's test).}
\scalebox{0.7}{
\begin{tabular}{ccc|ccc }
\hline
Model & Penalty & Flagged Changepoints &$\hat{\sigma}^2$ & Log-likelihood  & Penalized Score  \\ \hline
\multirow{2}{*}{Trend shifts+AR(1)} & BIC &1987 &0.305 &-205.44 &449.51 \\
& MDL &1987 &0.305 &-205.44 & 450.70 \\ 
\hline
\multirow{2}{*}{Trend shifts+WN} & BIC & { 1987} &{  0.308}  &{ -206.13} & \textbf{445.36} \\
& MDL &{ 1987}  &{0.308} & { -206.13}  & \textbf{446.55}\\ 
\hline
\multirow{2}{*}{Fixed slope mean shift+AR(1)} & BIC & 1990 &0.306 &-208.06 &449.23 \\
& MDL  &1990 &0.306 &-208.06  &452.51\\
\hline
Joinpin & BIC & - & 0.308* & -220.72 & 452.47\\
\hline
Long-memory+AR(1) & BIC & - & 0.333 & -217.01 & 450.57\\
\hline
Long-memory & BIC & - & 0.340 & -219.41 & 449.85\\
\hline
\end{tabular}
\label{tab:Truncatedmod_fitting_result}
}
\end{table}

\begin{table}
\centering
\caption{BIC posterior probabilities for models fitted to the full and truncated CET series}
\begin{tabular}{ >{\centering\arraybackslash}p{6cm}  >{\centering\arraybackslash}p{3cm}  >{\centering\arraybackslash}p{3cm}}
\hline
Model    &   Full  & Truncated \\
[0.3em]
\hline
Trend shifts + AR(1) & 0.11 & 0.08  \\
[0.3em]
Trend shifts + WN &  0.64  & 0.68\\
[0.3em]
Fixed slope +mean shifts+AR(1) &  0.05  & 0.10\\
[0.3em]
Joinpin &  0.12 & 0.02  \\
[0.3em]
Long-memory+AR(1) &  0.03 & 0.05 \\
[0.3em]
Long-memory &  0.05 & 0.07\\
\hline
\end{tabular}
\label{tab:BICweights}
\end{table}

\begin{table}
\centering
\caption{Parameter estimates of the best fitting model: trend shifts with white noise errors}
\label{tab:best_fitting_mod}
\begin{subtable}{1\textwidth}
\centering
\begin{tabular}{ >{\centering\arraybackslash}p{3cm}  >{\centering\arraybackslash}p{3cm}  >{\centering\arraybackslash}p{3cm}}
\hline
Segment     & Slope ($^\circ$C/yr) \\
[0.3em]
\hline
1659-1699       &-0.027  \\
[0.3em]
1700-1738        &0.026\\
[0.3em]
1739-1987        &0.002\\
[0.3em]
1988-2020       &0.011\\
\hline
\end{tabular}
\caption{Full CET}
\label{tab:best_mod_fullcet}
\end{subtable}

\vspace{1cm}

\begin{subtable}{1\textwidth}
\centering
\begin{tabular}{ >{\centering\arraybackslash}p{3cm}  >{\centering\arraybackslash}p{3cm}  >{\centering\arraybackslash}p{3cm}}
\hline
Segment        & Slope ($^\circ$C/yr) \\
[0.3em]
\hline
1772-1986      &0.002   \\
[0.3em]
1987-2020     &0.016\\
\hline
\end{tabular}
\caption{Truncated CET}
\label{tab:best_mod_trunccet}
\end{subtable}
\end{table}

%%%%%%%%%%%%%%%%%%%%%%%%%%%%%%%%%%%%%%%%%%%%%%%%%%%%%%%%%%%%%%%%%%%%%
% ACKNOWLEDGMENTS
%%%%%%%%%%%%%%%%%%%%%%%%%%%%%%%%%%%%%%%%%%%%%%%%%%%%%%%%%%%%%%%%%%%%%
\section*{Acknowledgement}
Rebecca Killick gratefully acknowledges funding from EP/R01860X/1 and NE/T006102/1.  Robert Lund and Xueheng Shi thank funding from NSF DMS-2113592.

%%%%%%%%%%%%%%%%%%%%%%%%%%%%%%%%%%%%%%%%%%%%%%%%%%%%%%%%%%%%%%%%%%%%%
% DATA AVAILABILITY STATEMENT
%%%%%%%%%%%%%%%%%%%%%%%%%%%%%%%%%%%%%%%%%%%%%%%%%%%%%%%%%%%%%%%%%%%%%
% 
%
\section*{Data statement}
The Central England data used in this study is available at \url{https://www.metoffice.gov.uk/hadobs/hadcet/}.  We used the annual records from 1659-2020.

\bibliographystyle{unsrt}

\end{document}